\documentstyle[11pt,paspconf,epsfig]{article}

\newcommand{\iras}{{\sl IRAS\/}}

\newcommand{\avg}[1]{{\langle{#1}\rangle}}

\def\simless{\mathbin{\lower 3pt\hbox
	{$\,\rlap{\raise 5pt\hbox{$\char'074$}}\mathchar"7218\,$}}} 
\def\simgreat{\mathbin{\lower 3pt\hbox
	{$\,\rlap{\raise 5pt\hbox{$\char'076$}}\mathchar"7218\,$}}} 

\begin{document}

\title{Nonlinear and Stochastic Morphological Segregation}

\author{Michael Blanton}
\affil{Department of Astrophysics, Princeton University, Princeton, NJ
08544-1001}




\begin{abstract}
I perform a joint counts-in-cells analysis of galaxies of different
spectral types using the Las Campanas Redshift Survey (LCRS). Using a
maximum-likelihood technique to fit for the relationship between the
density fields of early- and late-type galaxies, I find a relative
linear bias of $b=0.76\pm 0.02$. This technique can probe the
nonlinearity and stochasticity of the relationship as well. However,
the degree to which nonlinear and stochastic fits improve upon the
linear fit turns out to depend on the redshift range in question. In
particular, there seems to be a systematic difference between the
high- and low-redshift halves of the data (respectively, further than
and closer than $cz\approx 36,000$ km/s); all of the signal of
stochasticity and nonlinearity comes from the low-redshift
portion. Analysis of mock catalogs shows that the peculiar geometry
and variable flux limits of the LCRS do not cause this effect. I
speculate that the central surface brightness selection criteria of
the LCRS may be responsible.
\end{abstract}




\section{Introduction}

Different types of galaxies have different large-scale density fields
(\cite{hubble36ap}; \cite{oemler74ap}).  Various authors have compared
elliptical and spiral galaxies, finding that the fluctuation amplitude
of ellipticals is stronger than that of spirals by a factor of
1.3--1.5 (\cite{santiago92ap}; \cite{loveday96ap}; \cite{hermit96ap};
\cite{guzzo97ap}).  Similarly, a comparison of the galaxy distribution
in the \iras\ redshift survey (\cite{strauss92bp}) with those in the
Center for Astrophysics Redshift Survey (CfA; \cite{huchra83ap}) and
in the Optical Redshift Survey (ORS; \cite{santiago95ap}), shows that
optically-selected galaxies are clustered more strongly than
infrared-selected galaxies by a similar factor (\cite{davis88ap};
\cite{babul90ap}; \cite{strauss92ap}).

All of these statistics, however, only compare the {\it amplitudes} of
the fluctuations of each galaxy type. It may be that the pattern of
fluctuations is different for different galaxy types, as well. That
is, that the local density of early-type galaxies is not
deterministically related to the local density of late-type galaxies;
other variables may affect the relationship. One way to quantify this
is by using the correlation coefficient $r_{el}\equiv\langle\delta_e
\delta_l\rangle/\sigma_e\sigma_l$, where $\delta$ refers to the galaxy
overdensity and $\sigma^2\equiv\avg{\delta^2}$ is its variance.
$r_{el}=1$ if the fields are perfectly
correlated, and $r_{el}=0$ if they are completely uncorrelated. In
large-volume hydrodynamic simulations of galaxy formation,
\cite{blanton99a} have found a low correlation coefficient
between the density fields of old and young galaxies, around
$r_{el}\sim 0.5$. The differences arise in large part from the fact
that the dependence of each galaxy density field on temperature is
different; young galaxies are not found in high-temperature regions in
the simulations.

Motivated by these theoretical results, in this proceeding I perform a
point-by-point comparison of the density fields of different galaxy
types, using a counts-in-cells analysis of the Las Campanas Redshift
Survey (LCRS; \cite{shectman96ap}). The maximum-likelihood method I
use is superior to the comparision of second moments alone
(\cite{tegmark99ap}), as I will demonstrate below, and can probe the
nonlinearity and stochasticity of the relationship between different
galaxy types. I describe my method in Section 2, describe the nature
of the data in Section 3, describe my results in Section 4, and show
results from mock catalogs in Section 5. The results presented at the
meeting have been amended; in particular, the stochasticity detected
between the density fields turns out to depend critically on redshift,
as explained more completely below. I conclude in Section 6.

\section{Method}
\label{c5_method}

I am interested in $f(\delta_e, \delta_l)$, the joint distribution of
the overdensities $\delta_e$ of early-type galaxies and $\delta_l$ of
late-type galaxies. \cite{dekel99a} show how such a joint distribution
determines the relationship between the correlation functions, the
power spectra, and higher-order moments of two density fields. It is
possible to constrain the properties of $f(\delta_e, \delta_l)$ using
these quantities. However, a more direct approach is to analyze the
related joint probability distribution $P(N_e, N_l)$ of finding $N_e$
early-type and $N_l$ late-type galaxies in a single cell of size
$R$. After all, this latter probability is simply $f(\delta_e,
\delta_l)$ (where the fields are smoothed over the volume of a cell)
convolved with Poisson distributions. If one notes that
$f(\delta_e,\delta_l) = f(\delta_l|\delta_e) f(\delta_e)$, one can
write
\begin{eqnarray}
\label{c5_prob2}
&P(N_e, N_l | \alpha)& = \int d\delta_e 
\frac{{\bar N}_e^{N_e} (1+\delta_e)^{N_e}}{N_e!} e^{-{\bar
N_e}(1+\delta_e)} f(\delta_e) \cr
&&\times \int d\delta_l 
\frac{{\bar N}_l^{N_l} (1+\delta_l)^{N_l}}{N_l!} e^{-{\bar
N_l}(1+\delta_l)}
f(\delta_l| \delta_e),
\end{eqnarray}
where ${\bar N}_e$ and ${\bar N}_l$ are the average number of galaxies
of each type expected in a cell of a given volume (and given selection
criteria), and $\alpha$ represents the parameters of $f(\delta_e,
\delta_l)$.  One can then fit for $\alpha$ by minimizing the quantity
${\mathcal L} \equiv -2 \sum_i \ln P(N_{e,i}, N_{l,i}|\alpha)$.
In practice, I first fit for
the parameters of the density distribution $f(\delta_e)$; then I fix
the parameters of $f(\delta_e)$ and  fit separately for those of
$f(\delta_l|\delta_e)$. I calculate error bars using the well-known
bootstrap method.

I parameterize $f(\delta_l, \delta_e)$ by first specifying the density
distribution function $f(\delta_e)$ and then the bias relation
$f(\delta_l|\delta_e)$.  Here, I will use a log-normal model for the
density distribution function:
\begin{equation}
\label{c5_lnden}
f(\delta_e) = \frac{1}{\sqrt{2\pi}\sigma_e(1+\delta_e)}
\exp\left[-x_e^2/2\sigma_e^2\right].
\end{equation}
where $x_e\equiv\ln(1+\delta_e)+\sigma_e^2/2$.  This model is a better fit
to the data than either a Gaussian or a first-order Edgeworth
expansion (\cite{bernardeau94ap}; \cite{juszkiewicz95ap};
\cite{kim98ap}); however, the results presented here do not depend
sensitively on my choice for $f(\delta_e)$.

The bias relation $f(\delta_l|\delta_e)$ can be either deterministic
or stochastic.  These terms are not meant to refer to underlying
physical principles but are simply meant to express whether knowing
the density of ellipticals tells you with certainty the density of
spirals, modulo Poisson noise. In the case of deterministic bias, the
joint density distribution can be expressed as
\begin{equation}
\label{c5_detbias}
f(\delta_l| \delta_e) = \delta^D(\delta_l - b(\delta_e))
\end{equation}
where $f(\delta_e)$ is the density distribution function of early-type
galaxies. Under this assumption, one can describe the models by the
function $b(\delta_e)$. 

The simplest model for $b(\delta_e)$ is linear bias:
\begin{equation}
\label{c5_linear}
b(\delta_e) = b_0+b_1 \delta_e,
\end{equation}
which can be trivially extended to quadratic bias:
\begin{equation}
\label{c5_quad}
b(\delta_e) = b_0+b_1 \delta_e + {b_2}\delta_e^2.
\end{equation}
Another possibility is ``broken'' bias, which is piece-wise
linear with one slope in overdense regions and another in underdense
regions:
\begin{equation}
\label{c5_broken}
b(\delta_e) = \left\{\begin{array}{ccc}
b_0+b_{1}\delta_e & {\mathrm{for}} & \delta_e<0 \cr
b_0+b_{2} \delta_e & {\mathrm{for}} & \delta_e>0 
\end{array}\right.
\end{equation}
I require that $\avg{\delta_l}=0$, because it is meant to represent
the overdensity of late-type galaxies. In practice, this requirement
sets $b_0$, which is therefore not treated as a free parameter in any
of the above expressions.

If variables other than the local density field are important in
determining where galaxies form, it may be that the different
formation processes of early-type and late-type galaxies cause scatter
in their relationship.  Thus, I examine models which incorporate
scatter by rewriting Equation (\ref{c5_detbias}), replacing the Dirac
delta-function with a Gaussian of finite width:
\begin{equation} 
\label{c5_stobias}
f(\delta_l|\delta_e) = \frac{1}{\sqrt{2\pi}\sigma_b} \exp\left[
\frac{(\delta_l- b(\delta_e))^2}{2\sigma_b^2}\right] 
\end{equation} 
where $b(\delta_e)$ and $f(\delta_e)$ are chosen as above.  In the
case that $f(\delta_e)$ is Gaussian, and $\sigma$ and $\sigma_b$ are
small, such that the limit $\delta\ge -1$ is not important,
$f(\delta_e, \delta_l)$ reduces to a bivariate Gaussian distribution,
and the standard correlation coefficient is related to $\sigma_b$ by
\begin{equation}
r = \sqrt{1-(\sigma_b/\sigma_l)^2}
\end{equation}
I will use Equation (\ref{c5_stobias}) to fit linear bias with
Gaussian scatter to the relationship between galaxy types.

\section{Galaxies in the LCRS}
\label{c5_data}

The LCRS (\cite{shectman96ap}) consists of $\sim$ 25,000 galaxies
with a median redshift of $z\sim 0.1$. Three long slices
($1.5^\circ\times80^\circ$) were surveyed in the North Galactic Cap,
and three in the South Galactic Cap. Within each hemisphere, the
slices had the same right-ascension limits but were separated by
several degrees in declination.  In addition to the flux limits, a cut
was applied on the central magnitude (the magnitude within the central
two pixels of the CCD images), to avoid putting fibers onto
galaxies unlikely to yield redshifts.  As in all redshift surveys,
this cut can affect the relationship between the luminosity function
and the selection function, since the selection is not purely based on
flux.

\cite{bromley98a} have used a spectral classification scheme to divide
the galaxies into six ``clans.''  For my purposes, I will split the
galaxies into just two groups: an early-type group consisting of clans
1 and 2 and a late-type group consisting of clans 3 through 6. I
place absolute magnitude limits on the early-type group of
$-22.5<M<-18.8$ and on the late-type group of $-22.0<M<-18.5$. This
procedure yields about 10,000 galaxies in each group.

The geometry of the LCRS complicates an attempt to perform a
counts-in-cells analysis on it.  I create 14 redshift shells, each
with an equal volume; thus, the shells at higher redshift have a
shorter radial extent.  In the angular dimension, I divide the survey
into cells which are 3 spectrograph fields on each side (each field is
about $1.5^\circ \times 1.5^\circ$). In the right ascension direction,
the fields are adjacent; in the declination direction, the fields from
the three slices in each Galactic hemisphere are combined.  This
procedure produces 518 cells total, each with a volume equivalent to a
15 $h^{-1}$ Mpc sphere, of about cubical dimensions at $z\sim 0.1$
(except for the gaps in the declination direction).  

\section{Results from the LCRS}
\label{c5_results}

\begin{figure}[t]
\plotfiddle{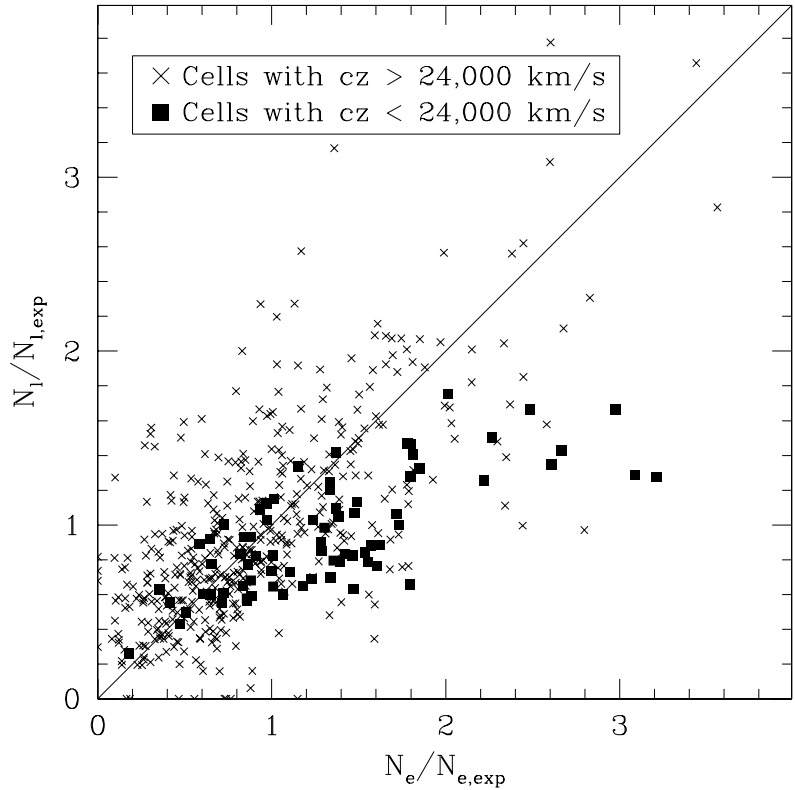}{3.8in}{0.}{60.}{60.}{-190in}{-120in}
\caption{\label{c5_cells} Joint overdensity distribution of early-type
galaxy {\it (x-axis)} and late-type galaxy {\it (y-axis)}
counts-in-cells for the LCRS. The low-redshift cells are shown as
solid squares and the high-redshift cells are shown as crosses. Note
that it appears as if the low-redshift cells are systematically less
biased than the high redshift cells, which accounts for the increased
stochasticity and nonlinearity when these cells are included. This low
bias could occur because the selection function for the late-type 
galaxies is overestimated at low redshifts.}
\end{figure}

I derive the luminosity and selection functions from the data and
calculate the expected counts in each cell for each galaxy type. The
distribution of $N/N_{\mathrm{exp}}$ is shown in Figure \ref{c5_cells}
for each galaxy type. The top quarter of Table \ref{c5_biasrd} shows
the results of fitting for the log-normal distribution and each bias
model.  Note first that for the linear fit $b_1=0.76\pm 0.02$; that
is, the late-type galaxies are underbiased with respect to the
early-type galaxies, to a degree which agrees with conventional
wisdom.  Some nonlinearity is detected in the quadratic bias case, at
rather high significance (6$\sigma$), in the sense that the slope of
the bias steepens at large densities ({\it i.e.} $b_2>0$). The broken
bias model also shows this effect, at somewhat lower significance.
Stochasticity is detected at about 10$\sigma$ significance, with
$\sigma_b=0.21\pm 0.02$.  This level of stochasticity corresponds to
$r\approx 0.87$; for comparison, the moments method of
\cite{tegmark99a} would estimate $r$ for this counts-in-cells
distribution to be $r = 0.73 \pm 0.01$.  Apparently a considerable
amount of the ``stochasticity'' measured by the moments method is due
to an inadequate characterization of the distribution of the
densities, most likely because the method does not account for the
non-Gaussianity of the Poisson distribution at low $N$ or for the
lower limit of $\delta\ge -1$.

The likelihoods relative to the linear fit are also shown in Table
\ref{c5_biasrd} for all the other models, and indicate that the
stochastic linear bias is clearly the best model I have
considered. I have also compared each of the two-parameter fits to
the linear fit by using a likelihood ratio test, whose results I show
in the last column as the probability $P_{\mathrm{random}}$ of getting
the likelihood difference between the linear model and the
two-parameter model by chance. Note that since I only ran 200
realizations for each estimate, there is a lower limit on
$P_{\mathrm{random}}$ of 0.005. From these results, it is clear that
I detect nonlinearity at a statistically significant level, and
stochasticity at an extremely significant level. 

As part of an effort to test for systematic effects, I split the cells
into a high-redshift half and a low redshift half. Results for each
half are shown in the middle two quarters of Table \ref{c5_biasrd}. I
find that the low redshift half retains the large amount of
nonlinearity and stochasticity, while the high redshift half has no
detectable signal. I can be more precise and eliminate the two
innermost rings of cells, and fit to the rest, as shown in the bottom
quarter of Table \ref{c5_biasrd}. This set of cells shows no
nonlinearity, and a much reduced stochasticity, indicating that most
of the signal for stochastic bias was coming from the two innermost
rings of cells; tests show that this effect is {\it not} due to the
better signal-to-noise in the inner rings. Again I can express the
stochasticity in terms of the correlation coefficient $r\approx 0.95$;
the moments method of \cite{tegmark99a} obtains $r = 0.93 \pm 0.03$
for this set of cells.

In order to understand this effect, reconsider Figure
\ref{c5_cells}, where I have marked the cells in the inner two
rings with square boxes.  It is clear that these cells have a smaller
bias compared to the other cells. This probably
indicates that the selection function is overestimated for
late-type galaxies at low redshifts.  I have performed the same
analyses using selection functions based on the luminosity functions
of \cite{bromley98a}, and find the same effect. 

\begin{table}[t]
\caption{\label{c5_biasrd} Properties of PDF and bias fits to
LCRS joint counts-in-cells of early and late type galaxies, for
different redshift selections.}
\begin{center}\scriptsize
\begin{tabular}{rrrrrrrr}
$cz$ Range (km/s)& $\sigma_e$ & $\sigma_l$ & Bias Model & $b_1$ 
& $b_2$ or $\sigma_b$ & ${\mathcal L}$ & $P_{\mathrm{random}}$\cr
\cr
\tableline
\tableline
\cr
$10,000$--$46,000$ & $ 0.54 \pm  0.02$ & $ 0.41 \pm  0.02$ & 
Linear & $ 0.76 \pm  0.02$ & --- &   0.0 & N/A \cr
& & & Quadratic & $ 0.73 \pm  0.03$ & $ 0.18 \pm  0.03$ & $-13.1$ & $< 0.005$ \cr
& & & Broken & $ 0.64 \pm  0.05$ & $ 0.89 \pm  0.05$ &  $-5.2$ & 0.020
\cr 
& & & Stochastic & $ 0.63 \pm  0.05$ & $ 0.21 \pm  0.02$ & $-106.0$ & $<0.005$ 
\cr
$10,000$--$36,000$ & $ 0.50 \pm  0.03$ & $ 0.39 \pm  0.02$ & 
Linear & $ 0.75 \pm  0.03$ & --- &   0.0 & --- \cr
& & & Quadratic & $ 0.73 \pm  0.04$ & $ 0.19 \pm  0.05$ &  $-8.2$ &
0.005 \cr 
& & & Broken & $ 0.66 \pm  0.05$ & $ 0.85 \pm  0.06$ &  $-1.8$ & 0.165
\cr 
& & & Stochastic & $ 0.55 \pm  0.06$ & $ 0.22 \pm  0.02$ & $-119.7$ &
$<0.005$ \cr
$36,000$--$46,000$ & $ 0.61 \pm  0.03$ & $ 0.46 \pm  0.03$ &
Linear & $ 0.79 \pm  0.05$ & --- &   0.0 & --- \cr
& & & Quadratic & $ 0.81 \pm  0.06$ & $-0.06 \pm  0.05$ &  $-0.6$ &
0.420 \cr 
& & & Broken & $ 0.80 \pm  0.11$ & $ 0.78 \pm  0.08$ &  $-0.0$ & 0.955
\cr 
& & & Stochastic & $ 0.78 \pm  0.04$ & $ 0.07 \pm  0.05$ &  $-0.2$ & 0.370 \cr
$24,000$--$46,000$ & $ 0.56 \pm  0.03$ & $ 0.43 \pm  0.03$ &
Linear & $ 0.81 \pm  0.03$ & --- &   0.0 & --- \cr
& & & Quadratic & $ 0.81 \pm  0.04$ & $ 0.00 \pm  0.03$ &  $-0.0$ &
0.990 \cr 
& & & Broken & $ 0.82 \pm  0.07$ & $ 0.80 \pm  0.05$ &  $-0.0$ & 0.920 \cr
& & & Stochastic & $ 0.77 \pm  0.04$ & $ 0.13 \pm  0.03$ &  $-8.9$ &
$<0.005$ \cr 
\end{tabular}
\end{center}
\end{table}

\section{Results from Mock Catalogs}
\label{c5_mock}

Because of the peculiar geometry and selection effects of the LCRS it
is necessary to test these results against mock catalogs where I have
simulated the observational effects inherent in the survey.
I would also like to understand whether some of the
systematic trends with redshift found in the last section can be
explained by observational effects.  I run
particle-mesh simulations of a 300 $h^{-1}$ Mpc box using $256^3$
particles and $512^3$ grid cells.  I use the flat model with
$\Omega_m=0.4$ and $\Omega_\Lambda=0.6$. To select the late-type
galaxies, I simply pick dark matter particles at random. To select the
early-type galaxies, I smooth the density field with a 3 $h^{-1}$ Mpc
Gaussian filter, and apply a threshhold of $\delta_{c,e}=0.25$; every
dark matter particle above the threshold has an equal probability of
becoming an early-type galaxy.

To create realistic mock catalogs, I pick a random particle in the
simulation to represent the observer.  I then ``observe'' the galaxies
in the simulation box, using the angular and photometric limits of the
LCRS, as well as the number of fibers available in each field
(\cite{shectman96ap}).  Furthermore, there is a probability of failing
to observe the galaxy which is a function of its magnitude, determined
from the data. In the real LCRS, fibers could not be placed more
closely than $55''$; I implement that restriction in the mock
catalogs as well.  Finally, I include the appropriate magnitude and redshift 
errors.

As a benchmark, I take the simulation and divide it into cubic cells
about 25 $h^{-1}$ Mpc on a side. I subsample the galaxies such that
there are about 20--30 galaxies of each type in each cell. I refer to
this sample as the volume-limited catalog. It is free of all of the
selection effects associated with the real survey, as well as
redshift-space distortions.  The cells are equivalent in volume to the
cells described in Section 3. I will evaluate the degree
to which the selection effects affect the results by comparing
realistic mock catalogs to the results for these volume-limited cells.

The results of this comparison are listed in Table
\ref{c5_mock2data}. There is a significant difference between the
variances measured for the volume-limited catalog and for the
realistic mock catalog. I believe this is simply due to the fact that
the non-cubical shapes of the cells in the LCRS allow the cells to
probe effectively larger scales, thus driving down the variances.
Tests show that the effect is not due to the incomplete sampling or
due to fiber collisions.  The important result, however, is that the
bias is qualitatively the same (though with some quantitative
differences). In particular, there is no increased $\sigma_b$ for the
mock catalogs; the ratio $\sigma_b/\sigma_l$ {\it is} about 50\%
higher for the mock catalogs than the volume-limited catalogs, but
this level is still low enough for us to conclude that the
stochasticity I observe in the real catalog is not caused by any of
the observational effects modeled here.

\begin{table}[t]
\caption{\label{c5_mock2data} Properties of PDF and bias fits to
realistic mock catalogs and volume-limited mock catalogs.}
\begin{center}\scriptsize
\begin{tabular}{rrrrrrr}
Catalog Type & $\sigma_e$ & $\sigma_l$ & Bias Model & $b_1$ 
& $b_2$ or $\sigma_b$ & ${\mathcal L}$\cr
\cr
\tableline
\tableline
\cr
Volume-limited & $ 0.85 \pm  0.02$ & $ 0.54 \pm  0.01$ &
Linear & $ 0.72 \pm  0.01$ & --- &   0.0 \cr
& & & Quadratic & $ 0.74 \pm  0.01$ & $-0.01 \pm  0.00$ &  $-5.2$ \cr
& & & Broken & $ 0.67 \pm  0.01$ & $ 0.79 \pm  0.02$ & $-14.3$ \cr
& & & Stochastic & $ 0.72 \pm  0.01$ & $ 0.06 \pm  0.01$ &  $-8.8$ \cr
Flux-limited & $ 0.66 \pm  0.03$ & $ 0.46 \pm  0.02$ &
Linear & $ 0.76 \pm  0.02$ & --- &   0.0 \cr
& & & Quadratic & $ 0.77 \pm  0.02$ & $-0.05 \pm  0.01$ &  $-2.8$ \cr
& & & Broken & $ 0.70 \pm  0.04$ & $ 0.82 \pm  0.05$ &  $-2.0$ \cr
& & & Stochastic & $ 0.74 \pm  0.03$ & $ 0.07 \pm  0.02$ &  $-2.7$ \cr
\end{tabular}
\end{center}
\end{table}

\section{Conclusions}

I have presented a powerful method of analyzing the
nature of the relationship between different galaxy types. As a probe
of stochasticity it is superior to calculating second moments; in
addition, it can measure nonlinearity.  It would be both simpler to
implement and less susceptible to the problems of selection I have run
into here if it were used to analyze a volume-limited survey, which I
intend to do using the Optical Redshift Survey.

I have interpreted the redshift-dependence of the results as a problem
in the understanding of the selection effects of the survey. The
outstanding selection effect which I have {\it not} modeled yet in my
mock catalogs is the central magnitude cut.  I am in the process of
determining whether the effect on the selection function is strong
enough to cause the redshift-dependence found in Table
\ref{c5_biasrd}. Note, importantly, that the results of
\cite{tegmark99a} are susceptible to the same problems I have found
here, and should be interpreted accordingly.

In conclusion, the result of this work I take most seriously is the
bottom quarter of Table \ref{c5_biasrd}, which excludes the two
innermost rings, and indicates a bias which is linear, with perhaps
some mild scatter, and an amplitude of $b_1\approx 0.8$.

\acknowledgements

Thanks to Michael Strauss for advice on this work as well as comments
on the text. I am indebted to Benjamin Bromley, Daniel Koranyi, Huan
Lin, Max Tegmark, and Douglas Tucker for advice and for access to
source code.  This work was supported in part by the grants NAG5-2759,
NAG5-6034, AST93-18185, AST96-16901, and the Princeton University
Research Board.

%


\begin{thebibliography}
\baselineskip 12pt
\bibitem[Babul \& Postman~(1990)]{babul90a} 
Babul, A., \& Postman, M. 1990, \apj, 359, 280
\vspace{-6pt}
\bibitem[Bernardeau \& Kofman (1994)]{bernardeau94a}
Bernardeau, F., \& Kofman, L. 1995, \apj, 443, 479
\vspace{-6pt}
\bibitem[Blanton {\it et al.}~(1999)]{blanton99a} 
Blanton, M., Cen, R., Ostriker, J.~P., \& Strauss, M.~A.~1999, 
in press
\vspace{-6pt}
\bibitem[Bromley {\it et al.}~(1998)]{bromley98a} 
Bromley, B.~C., Press, W.~H., Lin, H., \& Kirshner, R.~P.~1998,
\apj, 505, 25
\vspace{-6pt}
\bibitem[Davis~{\it et al.} (1988)]{davis88a}
Davis, M., Meiksin, A., Strauss, M. A., da Costa, N., \&
Yahil, A. 1988, \apj, 333, L9
\vspace{-6pt}
\bibitem[Dekel \& Lahav~(1999)]{dekel99a} 
Dekel, A.~\& Lahav, O.~1999, \apj, 520, 24
\vspace{-6pt}
\bibitem[Guzzo~{\it et al.} (1997)]{guzzo97a}
Guzzo, L., Strauss, M.~A., Fisher, K.~B., Giovanelli, R., \& Haynes,
M.~P. 1997, \apj, 489, 37
\vspace{-6pt}
\bibitem[Hermit~{\it et~al.}~(1996)]{hermit96a} 
Hermit, S., Santiago, B.~X., Lahav, O., Strauss, M.~A., Davis, M., 
Dressler, A., \& Huchra, J.~P.~1996, \mnras, {283}, 709
\vspace{-6pt}
\bibitem[Hubble~(1936)]{hubble36a} 
Hubble, E. P. 1936, The Realm of the Nebulae (New Haven: Yale
University Press)
\vspace{-6pt}
\bibitem[Huchra {\it et al.}~(1983)]{huchra83a} 
Huchra, J., Davis, M., Latham, D., \& Tonry, J. 1983, \apjs, 52, 89
\vspace{-6pt}
\bibitem[Juszkiewicz {\it et al.}~(1995)]{juszkiewicz95a}
Juszkiewicz, R., Weinberg, D. H., Amsterdamski, P., Chodorowski,
M., \& Bouchet, F. R. 1995, \apj, 442, 39 
\vspace{-6pt}
\bibitem[Kim \& Strauss~(1998)]{kim98a}
Kim, R.~S., \& Strauss, M.~A.~1998, \apj, 493, 39
\vspace{-6pt}
\bibitem[Loveday {\it et al.} (1996)]{loveday96a}
Loveday, J., Efstathiou, G., Maddox, S.~J., \& Peterson, B.~A. 1996,
\apj, 468, 1
\vspace{-18pt}
\bibitem[Oemler~(1974)]{oemler74a} 
Oemler, A.~1974, \apj, 194, 1
\vspace{-6pt}
\bibitem[Santiago \& Strauss~(1992)]{santiago92a} 
Santiago, B.~X.~\& Strauss, M.~A.~1992, \apj, {387}, 9
\vspace{-6pt}
\bibitem[Santiago {\it et al.}~(1995)]{santiago95a} 
Santiago, B.~X., Strauss, M.~A., Lahav, O., Davis, M., Dressler, A., 
\& Huchra, J.~P. 1995, \apj, 446, 457
\vspace{-6pt}
\bibitem[Shectman {\it et al.}~(1996)]{shectman96a}
Shectman, S.~A., Landy, S.~D., Oemler, A., Tucker, D.~L., Lin, H., 
Kirshner, R.~P., \& Schechter, P.~L. 1996, \apj, 470, 172 
\vspace{-6pt}
\bibitem[Strauss~{\it et~al.}~(1992a)]{strauss92a}
Strauss, M.~A., Davis, M., Yahil, A., \& Huchra J.~P. 1992a, \apj, {385}, 421
\vspace{-6pt}
\bibitem[Strauss~{\it et~al.}~(1992b)]{strauss92b}
Strauss, M.~A., Huchra, J.~P., Davis, M., Yahil, A., Fisher, K.~B., \&
Tonry, J. 1992b, \apjs, {83}, 29
\vspace{-6pt}
\bibitem[Tegmark \& Bromley~(1999)]{tegmark99a}
Tegmark, M.,~\& Bromley, B.~1999, \apj, 518, L69
\bibitem[Babul \& Postman~1990]{babul90ap}
\vspace{-1in}
\bibitem[Bernardeau \& Kofman 1994]{bernardeau94ap}
\vspace{-1in}
\bibitem[Blanton {\it et al.}~1999]{blanton99ap}
\vspace{-1in}
\bibitem[Bromley {\it et al.}~1998]{bromley98ap}
\vspace{-1in}
\bibitem[Davis~{\it et al.} 1988]{davis88ap}
\vspace{-1in}
\bibitem[Dekel \& Lahav~1999]{dekel99ap}
\vspace{-1in}
\bibitem[Guzzo~{\it et al.} 1997]{guzzo97ap}
\vspace{-1in}
\bibitem[Hermit~{\it et~al.}~1996]{hermit96ap}
\vspace{-1in}
\bibitem[Hubble~1936]{hubble36ap}
\vspace{-1in}
\bibitem[Huchra {\it et al.}~1983]{huchra83ap}
\vspace{-1in}
\bibitem[Juszkiewicz {\it et al.}~1995]{juszkiewicz95ap}
\vspace{-1in}
\bibitem[Kim \& Strauss~1998]{kim98ap}
\vspace{-1in}
\bibitem[Loveday {\it et al.} 1996]{loveday96ap}
\vspace{-1in}
\bibitem[Oemler~1974]{oemler74ap}
\vspace{-1in}
\bibitem[Santiago \& Strauss~1992]{santiago92ap}
\vspace{-1in}
\bibitem[Santiago {\it et al.}~1995]{santiago95ap}
\vspace{-1in}
\bibitem[Shectman {\it et al.}~1996]{shectman96ap}
\vspace{-1in}
\bibitem[Strauss~{\it et~al.}~1992a]{strauss92ap}
\vspace{-1in}
\bibitem[Strauss~{\it et~al.}~1992b]{strauss92bp}
\vspace{-1in}
\bibitem[Tegmark \& Bromley~1999]{tegmark99ap}
\vspace{-1in}
\end{thebibliography}
\end{document}